\documentclass[aps,prc,twocolumn,superscriptaddress,showpacs,floatfix,nofootinbib]{revtex4-1}
\usepackage{amsmath,graphicx}

\begin{document}

\title{Directed flow at mid-rapidity in event-by-event hydrodynamics}

\author{Fernando G. Gardim}
\author{Fr\'ed\'erique Grassi}
\author{Yogiro Hama}
\affiliation{
Instituto de F\'\i sica, Universidade de S\~ao Paulo, C.P. 66318, 05315-970, S\~ao Paulo-SP, Brazil}
\author{Matthew Luzum}
\affiliation{
CEA, IPhT, Institut de physique theorique de Saclay, F-91191
Gif-sur-Yvette, France}
\author{Jean-Yves Ollitrault}
\affiliation{
CNRS, URA2306, IPhT, Institut de physique theorique de Saclay, F-91191
Gif-sur-Yvette, France}
\date{\today}

\begin{abstract}
Fluctuations in the initial geometry of a nucleus-nucleus
collision have been recently shown to result in a new type of
directed flow ($v_1$) that, unlike the usual directed flow, is
also present at midrapidity. We compute this new $v_1$ versus
transverse momentum and centrality for Au-Au collisions at RHIC
using the hydrodynamic code NeXSPheRIO. We find that the event
plane of $v_1$ is correlated with the angle of the initial dipole
of the distribution, as predicted, though with a large dispersion.
It is uncorrelated with the reaction plane. Our results are in
excellent agreement with results inferred from STAR correlation
data.
\end{abstract}


\maketitle

\section{Introduction}

Analyses of correlations between particles emitted in
ultrarelativistic heavy-ion collisions at large relative rapidity
reveal azimuthal structure that can be interpreted as solely due
to collective flow~\cite{Luzum:2010sp}.
Event-by-event hydrodynamics~\cite{Hama:2004rr,Petersen:2008dd,Holopainen:2010gz,Schenke:2010rr,Werner:2010aa}
is a natural framework for studying collective flow.
In each event, particles are emitted independently according to some
momentum distribution determined by a fluid freeze out surface. The most general azimuthal distribution can be
written as a Fourier series with respect to the azimuthal angle $\phi$ of the
particle momentum:
\begin{equation}
\label{defvn}
2\pi\frac{dP}{d\phi}=1+2\sum_{n=1}^{+\infty}
v_n\cos(n(\phi-\Psi_{n})),
\end{equation}
where $v_n$
is the magnitude of anisotropic flow~\cite{Voloshin:1994mz}, and
$\Psi_{n}$ is the reference angle in harmonic $n$.
An equivalent definition is
\begin{equation}
\label{defvn2}
v_ne^{in\Psi_{n}}=\langle e^{in\phi}\rangle,
\end{equation}
where angular brackets denote an average with the probability law
$dP/d\phi$.
Once a convention is chosen for the
sign of $v_n$, these equations define
$\Psi_{n}$ up to $2\pi/n$.
Both $v_n$ and $\Psi_{n}$ may depend on the transverse momentum $p_t$
and the rapidity $y$.

Typically, the largest term in the series (\ref{defvn}) is elliptic flow,
$v_2$~\cite{Ackermann:2000tr},
and $\Psi_{2}$ is usually referred to as the event plane.
The only other harmonics measured so far at RHIC are $v_1$ and
$v_4$~\cite{Adams:2003zg}.
However, it has recently been pointed out that fluctuations in the initial
density profile generally result in different reference angles $\Psi_{n}$
for every $n$. In particular, one expects additional
triangular ($v_3$)~\cite{Alver:2010gr} and dipole
($v_1$)~\cite{Teaney:2010vd} components from fluctuations,
both uncorrelated with the event plane $\Psi_{2}$.
While triangular flow has already been studied in a hydrodynamic framework~\cite{Schenke:2010rr,Qin:2010pf,Alver:2010dn}, there has not yet been a prediction for this  new $v_1$ that arises from fluctuations.

The variation of directed flow with rapidity can be uniquely separated
into even and odd parts:
\begin{equation}
\label{evenodd}
v_1(y) e^{i\Psi_{1}(y)}=
v_1^{\rm even}(y)e^{i\Psi_{1}^{\rm even}(y)}+
v_1^{\rm odd}(y)e^{i\Psi_{1}^{\rm odd}(y)},
\end{equation}
where $v_1^{\rm even}(y)$, $\Psi_{1}^{\rm even}(y)$ and $\Psi_{1}^{\rm
  odd}(y)$ are even functions of $y$, while
$v_1^{\rm odd}(-y)=-v_1^{\rm odd}(y)$.
Usual directed
flow~\cite{Adams:2003zg,Alt:2003ab,Adams:2005ca,Abelev:2008jga}
is $v_1^{\rm odd}(y)$, and the
corresponding
$\Psi_{1}^{\rm odd}$ is correlated with the reaction plane and $\Psi_{2}$.
Teaney and Yan recently argued~\cite{Teaney:2010vd} that
fluctuations in the initial geometry create an even part $v_1^{\rm
  even}(y)$, which depends weakly on $y$.
This even part does not contribute to existing directed flow measurements, but can be isolated
experimentally~\cite{Luzum:2010fb} using an event-plane method where
the weights are independent of rapidity (assuming that the detector is
symmetric in rapidity).
$\Psi_{1}^{\rm even}(y)$ is predicted to have little
correlation with the reaction plane or $\Psi_2$, unlike $\Psi_{1}^{\rm odd}(y)$.  It is also
expected to have little dependence on $p_t$ or $y$.

There is no dedicated analysis of  $v_1^{\rm  even}(y)$ yet, but
indirect evidence has been obtained~\cite{Luzum:2010fb}  from recent
STAR correlation data~\cite{Agakishiev:2010ur}, and both the magnitude
and $p_t$ dependence of $v_1$ are in qualitative agreement with
theoretical expectations~\cite{Teaney:2010vd}.

We present the first quantitative predictions for $v_1^{\rm
  even}$ in Au-Au collisions at the top RHIC energy, using
the hydrodynamic code NeXSPheRIO~\cite{Hama:2004rr}.
Calculations of $v_1^{\rm odd}$ will be presented
separately~\cite{Vezzoni}.
NeXSPheRIO solves the equations of relativistic ideal hydrodynamics
using initial
conditions provided by the event generator
NeXus~\cite{Drescher:2000ha}.
Fluctuations in initial conditions are studied by
generating many NeXus events, and solving the equations of
ideal hydrodynamics independently for each event.
NeXSPheRIO provides a good description of rapidity and transverse
momentum spectra~\cite{Qian:2007ff},
and elliptic flow $v_2$~\cite{Andrade:2008xh}.
In addition, it reproduces the ridge observed in two-particle
correlations~\cite{Takahashi:2009na}, which is produced by
initial fluctuations, followed by collective flow.
Teaney and Yan's dipole asymmetry is created by fluctuations followed by
collective flow, much in the same way as the
ridge~\cite{Alver:2010gr}, and should therefore be present in NeXSPheRIO.

NeXSPheRIO does not include effects of shear viscosity, which was
recently implemented in event-by-event
hydrodynamics~\cite{Schenke:2010rr}.
While shear viscosity produces a sizable reduction of elliptic
flow~\cite{Luzum:2008cw,Shen:2010uy}, we expect its effect to be
smaller on $v_1$, following the general observation that damping is
larger for higher harmonics~\cite{Alver:2010dn}.

In Sec.~\ref{s:method}, we explain how $v_1^{\rm even}$ is
calculated. In Sec.~\ref{s:predictions}, we present predictions for
its $p_t$ dependence and centrality dependence in Au-Au collisions at
the top RHIC energy and compare with existing STAR data.
In Sec.~\ref{s:initial}, we study the correlation of
directed flow with the initial dipole asymmetry defined by
Teaney~\cite{Teaney:2010vd}. Our conclusions are presented in
Sec.~\ref{s:conclusions}.
We use $v_1$, $\Psi_{1}$ as a shorthand notation for $v_1^{\rm
  even}$, $\Psi_{1}^{\rm even}$  throughout this paper.

\section{Measuring $v_1$ in event-by-event hydrodynamics }
\label{s:method}

The code NeXSPheRIO emits particles at the end of the hydrodynamical
evolution using a Monte-Carlo generator, and one can analyze
events much in the same way as in an actual experiment.
Particles are emitted independently. The only nonflow correlation in
NeXSPheRIO is the correlation induced by resonance
decays~\cite{Borghini:2000cm}.
In principle, $v_1(p_t,y)$ and $\Psi_{1}(p_t,y)$ can be computed
directly from Eq.~(\ref{defvn2}) by averaging over particles emitted
from the Monte-Carlo generator.
The only practical limitation is computer time, which results in
finite statistical errors. In order to obtain reliable results,
we integrate $v_1$ over the pseudorapidity interval $-1<\eta<1$,
corresponding approximately to the acceptance of the time projection
chamber of the STAR experiment~\cite{Ackermann:1999kc}.
We estimate $\Psi_{1}$ by computing a weighted average $\Psi_{EP,1}$ inspired by the
event-plane method used in experimental analyses~\cite{Poskanzer:1998yz}:
\begin{eqnarray}
\label{eventplane}
Q\cos\Psi_{EP,1}\equiv\sum_{|\eta|<1} w_j \cos\phi_j\cr
Q\sin\Psi_{EP,1}\equiv\sum_{|\eta|<1} w_j \sin\phi_j
\end{eqnarray}
where the sum is over particles emitted in the pseudorapidity
interval $-1<\eta<1$, $\phi_j$ are the azimuthal angles of outgoing
particles, $Q\ge 0$,
and $w_j$ is a weight~\cite{Luzum:2010fb} depending on the particle
transverse momentum $p_t$:
\begin{equation}
\label{weight}
w(p_t)=p_t-\frac{\langle p_t^2\rangle}{\langle p_t\rangle},
\end{equation}
where angular brackets denote an average over particles of all
events in the centrality window and in the same pseudorapidity
interval $-1<\eta<1$.
This particular choice of the weight eliminates both
$v_1^{\rm odd}$ in Eq.~(\ref{evenodd}), and nonflow correlations from
momentum conservation (though the latter are not implemented in
this calculation).
We compute  $\langle p_t^2\rangle$ and $\langle p_t\rangle$ by
averaging over all events in the 0-40\% centrality window, where $p_t$
spectra depend little on centrality~\cite{Adler:2003cb}.
$w(p_t)$ changes sign at $p_t=\langle p_t^2\rangle/\langle
p_t\rangle\simeq 0.85$~GeV/c.

In order to achieve small statistical errors, we run the Monte-Carlo
generator many times for each event.
In practice, the summation in
Eq.~(\ref{eventplane}) runs over $7\times 10^5$ particles.
This ensures almost perfect reconstruction of $\Psi_{1}$ for most
events. We then measure
\begin{eqnarray}
\label{v1mc}
v_{1}(p_t)&\equiv&\left\langle\cos(\phi-\Psi_{EP,1})\right\rangle\cr
v_{1s}(p_t)&\equiv&\left\langle\sin(\phi-\Psi_{EP,1})\right\rangle
\end{eqnarray}
where brackets denote an average over particles in a $p_t$ bin.
If $\Psi_{1}$ in Eq.~(\ref{defvn2}) is independent of $p_t$, then
it coincides with $\Psi_{EP,1}$, and $v_{1s}(p_t)$ vanishes identically.
\begin{figure}[ht]
\includegraphics[width=\linewidth]{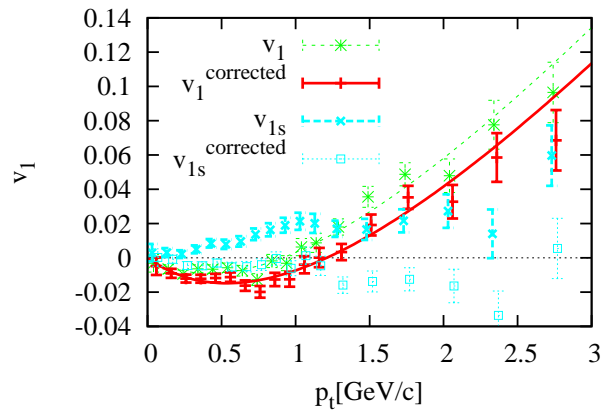}
\caption{(Color online)
Hadron $v_1(p_t)$ and $v_{1s}(p_t)$ in $|\eta|<1$ for a single
NeXSPheRIO event with impact parameter $b=1$~fm.
Results are shown with and without correcting for the net transverse
momentum (see text). Vertical bars are statistical errors. }
\label{fig:event1}
\end{figure}

Fig.~\ref{fig:event1} displays $v_1(p_t)$ and $v_{1s}(p_t)$ for the
first event we analyzed.
$v_1(p_t)$ has roughly the expected behavior: it
vanishes linearly at low $p_t$~\cite{Ollitrault:1997vz}, changes sign
around 1~GeV~\cite{Teaney:2010vd}, and increases
linearly at high $p_t$~\cite{Borghini:2005kd}.
The solid line is a 3-parameter rational fit with the same properties:
\begin{equation}
\label{fit}
v_1(p_t)=\frac{a p_t (p_t-b)}{p_t+c}.
\end{equation}
 $v_{1s}(p_t)$ is not compatible with 0: it is in fact positive for
 all $p_t$, which means that there is a small nonzero net transverse
 momentum in the direction perpendicular to $\Psi_{EP,1}$.
Averages in Eq.~(\ref{v1mc}) are taken over the pseudorapidity
interval $-1<\eta<1$.

This net transverse momentum could come from several sources.
In principle, net transverse momentum can be generated by the initial dynamics, if a  corresponding transverse momentum is transferred to ``spectator'' nucleons.  An additional net transverse momentum can be generated in the pseudorapidity interval used here if, during the evolution of the system, net momentum is transferred to particles that end up outside this range of pseudorapidity.
In addition, however, there can be spurious momentum generated in
NeXSPheRIO during the thermalization stage, as an unphysical side affect of the transition from the microscopic model NeXuS to hydrodynamics.  A detailed explanation and the alternative procedure that exactly conserves conservation of energy and momentum is presented in Appendix~\ref{s:app}, though its implementation and a thorough investigation of energy and momentum non-conservation in the standard scheme are left for future work.


We are not aware of any measurement of the net transverse momentum of
charged hadrons in a symmetric pseudorapidity interval. In particular,
there is no experimental evidence that it is larger than statistical
fluctuations.
Since our $v_1$ results can be significantly modified by a net momentum, which is
yet unknown experimentally, it is useful to present another set of results with a different
(smaller) value of the net momentum.

A simple way to correct for the net transverse momentum is by boosting the system
in the transverse direction, and by adjusting the boost velocity so
that the net momentum vanishes in the new frame.
This is  difficult in practice, however,  because the ``system'' here
is the set of particles in the pseudorapidity interval $-1<\eta<1$,
and $\eta$ is not invariant under a transverse boost.
We adopt the following approximate procedure: we define the boost
velocity by $\vec v\equiv \sum_{|\eta|<1}\vec p/\sum_{|\eta|<1}E$.
We then boost all the particles, select particles with $|\eta|<1$ in
the new frame, and compute $v_1$ for these particles.
Since this set of particles differs from the original set, the net
transverse momentum is not strictly zero in the new frame, but
it should be significantly smaller.
Results after correction are shown in Fig.~\ref{fig:event1}.
$v_{1s}(p_t)$ is smaller in absolute magnitude, but still nonzero,
because the correction is approximate.

The next step is to average over events.
This is done in the same way as would be done in an experiment.
The scalar-product method~\cite{Adler:2002pu} provides a simple
prescription for doing the average, and it is known to give the same
results as the event-plane method if the event-plane resolution is not
too large~\cite{Ollitrault:2009ie}.
Eq.~(7) of \cite{Adler:2002pu} gives:
\begin{equation}
\label{spmethod}
v_1\{EP\}=\frac{\langle v_1 Q\rangle}{\sqrt{\langle Q^2\rangle}},
\end{equation}
where angular brackets denote average over particles in the
numerator, and over events in the denominator. Due to the (almost perfect)
event-plane resolution, there is no need to use subevents: the flow
vector of one subevent is approximately half of the total flow vector,
$Q_a\simeq Q_b\simeq Q/2$.

Results presented in this paper are obtained from a set of 120
different NeXSPheRIO events, each corresponding to a different initial
geometry. Each event is a Au-Au collision at 200 GeV per nucleon in
the 0-60\% centrality range. Events are uniformly distributed in
centrality, and centrality is determined according to the number of
participant nucleons.
Compared with an actual experiment, we work with a limited number of
events, but we generate a very large multiplicity in every event, so
that our determination of $v_1$ is not limited by statistical
fluctuations.
In practice, 75\% of the events that we use have a resolution
parameter~\cite{Ollitrault:1997vz} $\chi\equiv Q/\sqrt{\sum (w_j)^2}$
larger than 3, implying
an event plane resolution $\langle\cos\Delta\phi\rangle$ larger than
0.97. Since $\chi$ scales like $Q$, events with the lowest resolution
have a negligible contribution to $v_1\{EP\}$.

\section{Predictions for $v_1$ at RHIC}
\label{s:predictions}

\begin{figure}[ht]
\includegraphics[width=\linewidth]{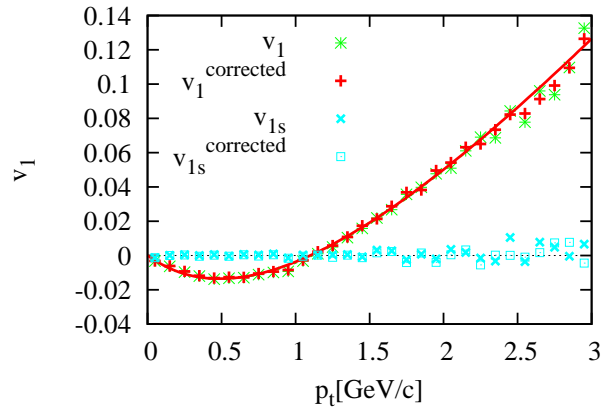}
\caption{(Color online)
$v_1(p_t)$ and $v_{1s}(p_t)$ of hadrons in $|\eta|<1$
averaged over events in the $0-40\%$ centrality window.
As in Fig.~\ref{fig:event1}, results are shown with and without
correcting for the net transverse momentum.
Solid lines are fits using Eq.~(\ref{fit}). }
\label{fig:pt}
\end{figure}
Our results for $v_1(p_t)$ and $v_{1s}(p_t)$ in
the centrality bin $0-40\%$  are presented in Fig.~\ref{fig:pt},
with and without correcting for the net transverse momentum in each
event.
After averaging over events, $v_{1s}(p_t)$ is much smaller in absolute
magnitude than $v_1(p_t)$.
In fact, $v_{1s}(p_t)$ should be identically zero by parity symmetry.
Any difference can be attributed to statistical uncertainty due to the
finite number of events.
Results in this centrality range are very similar with or without correction for the net
transverse momentum.
The $p_t$ dependence of $v_1$ is in qualitative agreement with theoretical
expectations~\cite{Teaney:2010vd}: it is linear and negative at low $p_t$,
reaches a minimum between $-0.01$ and $-0.02$, then increases and
crosses 0 at $p_t\simeq 1.1$~GeV/c.


\begin{figure}[ht]
\includegraphics[width=\linewidth]{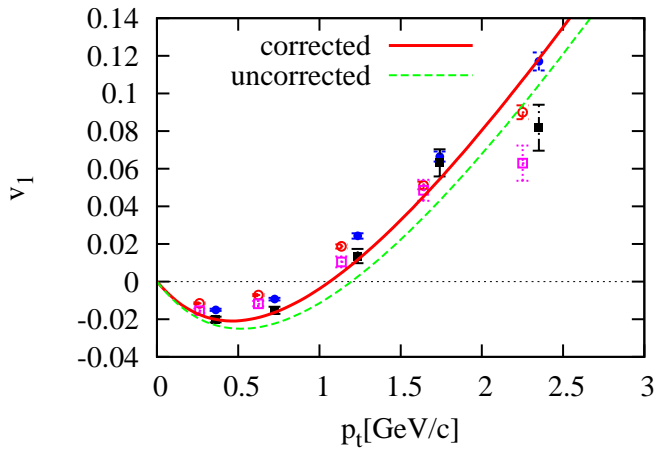}
\caption{(Color online)
$v_1(p_t)$ of hadrons in $|\eta|<1$ in Au-Au collisions
(20-60\% centrality) inferred from STAR correlation
data~\cite{Agakishiev:2010ur}. Different sets of points correspond to
different trigger particles and different assumptions concerning the
$v_1$ of trigger particles (see \cite{Luzum:2010fb}).
Solid curve: our hydro calculation
averaged over events in the same centrality window, corrected
for net transverse momentum (see text) and fit using
Eq.~(\ref{fit}).  The dashed curve represents the uncorrected value.}
\label{fig:data}
\end{figure}
Fig.~\ref{fig:data} compares our hydro calculation for the centrality
range 20-60\% to values of $v_1$~\cite{Luzum:2010fb} inferred from
STAR correlation data in the same centrality
range~\cite{Agakishiev:2010ur}.
As in Fig.~\ref{fig:pt}, results are shown with and without
correcting for the net transverse momentum in each event.
For this centrality range, $v_1(p_t)$ is slightly larger when the correction is made, resulting
in excellent agreement with data in the range $0<p_t<2$~GeV/c.

\begin{figure}[ht]
\includegraphics[width=\linewidth]{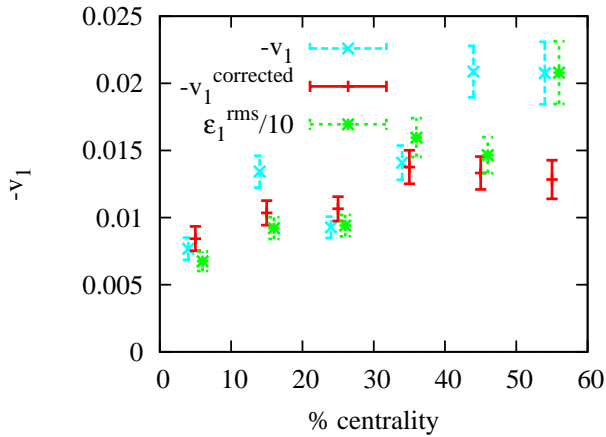}
\caption{(Color online)
$|v_1|=-v_1$ of hadrons with $0<p_t<1$, $|\eta|<1$ versus
centrality.
Dashed lines: raw  NeXSPheRIO results. Solid lines: after correcting
for net transverse momentum (see text). Dotted lines:
Rms dipole asymmetry  (scaled down by a factor 10)
versus centrality.
Error bars on $v_1$ are obtained by assuming that the relative
statistical errors on $v_1$ and $\varepsilon_1^{\rm rms}$ are identical.}
\label{fig:centrality}
\end{figure}
Fig.~\ref{fig:centrality} displays the centrality dependence of $-v_1$,
averaged over hadrons with $p_t<1$~GeV/c, $|\eta|<1$.
Results are smoother when  NeXSPheRIO results are corrected for the net
transverse momentum.
$-v_1$ increases mildly with centrality.
The reason is that $v_1$ is created by fluctuations, which are larger
for more peripheral collisions.
The centrality dependence of $v_1$ is comparable to that of
$v_3$~\cite{Alver:2010dn}.

\section{Correlation with the initial dipole asymmetry}
\label{s:initial}

Teaney and Yan~\cite{Teaney:2010vd} have shown that fluctuations
in the initial geometry of a nucleus-nucleus collision are expected to
create this new type of directed flow.
Fluctuations break the symmetry of the initial density profile, and as
a result there is, in general, one direction where the profile is
steepest.  This
effect can be quantified as a dipole asymmetry in the initial
density~\cite{Teaney:2010vd}:
\begin{equation}
\label{defpsi1}
\varepsilon_1 e^{i\Phi_{1}}=-\frac{\langle r^3 e^{i\phi}\rangle}{\langle r^3\rangle}.
\end{equation}
where the averages in the right-hand side are taken over the initial
transverse energy density profile, and  $(r,\phi)$ is a polar
coordinate system around the center of the distribution, defined by
$\langle r e^{i\phi}\rangle=0$. If one chooses $\varepsilon_1$
to be positive, then $\Phi_{1}$ generally corresponds  to the steepest
direction for a smooth profile, and $\varepsilon_1$ is the  magnitude
of the dipole asymmetry. In general,  $\varepsilon_1$ differs from
0  --- even at midrapidity --- due to fluctuations.

For smooth initial conditions, one expects
$\Psi_{1}=\Phi_{1}$ in each event. One also expects
$v_1\propto \varepsilon_1$, in the same way as
elliptic flow is
proportional to the participant eccentricity~\cite{Alver:2006wh}.
Experimentally, $v_1$ will be extracted from a two-particle
correlation, which scales like $(v_1)^2$, so that the experimentally
measured $v_1$ should scale like the rms dipole asymmetry
$\varepsilon_1^{\rm rms}\equiv\sqrt{\langle
  (\varepsilon_1)^2\rangle}$~\cite{Miller:2003kd}.
Fig.~\ref{fig:centrality} displays the rms dipole asymmetry
versus centrality.
The values are similar to those obtained using a Monte-Carlo
Glauber calculation~\cite{Teaney:2010vd}. In particular,
$\varepsilon_1^{\rm rms}$ is in general larger for more peripheral collisions,
and scales approximately with the number of participants like $N_{\rm
  part}^{-1/2}$, in the same way as
$\varepsilon_3$~\cite{Alver:2010dn}.
The variation of $-v_1$ for peripheral collisions is milder than that
$\varepsilon_1^{\rm rms}$. This can be attributed to the early freeze-out in
peripheral collisions, which has an effect analogous to viscosity and
breaks $\varepsilon_1$ scaling.

\begin{figure}[ht]
\includegraphics[width=0.8\linewidth]{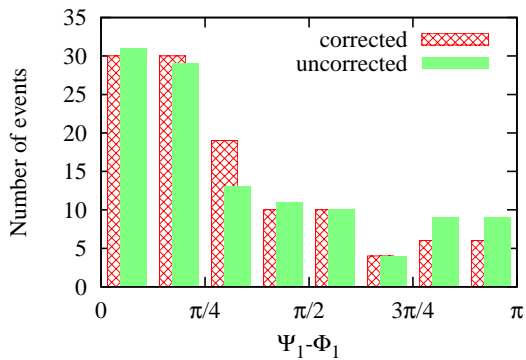}
\caption{(Color online)
Distribution of the relative angle between the event plane $\Psi_{1}$ and the
initial dipole $\Phi_{1}$ for the centrality range $0-60\%$.
As in previous figures, results are shown before and after correcting for
net transverse momentum. }
\label{fig:psi1}
\end{figure}
With bumpy initial conditions, there is no simple direct
correspondence between the initial geometry and the final
momentum distribution~\cite{Holopainen:2010gz}.
Fig.~\ref{fig:psi1} displays the correlation between the angle of
directed flow, $\Psi_{1}$, and the angle of the initial dipole
asymmetry, $\Phi_{1}$. The correlation is quite clear, which
confirms that the dipole asymmetry is a valid mechanism for generating
$v_1$. However, there is a large dispersion: some events
develop directed flow in a direction very different from the initial
dipole asymmetry.
A similar dispersion has also been observed for elliptic
flow between $\Psi_2$ and $\Phi_2$~\cite{Holopainen:2010gz} and
for triangular flow between $\Psi_3$ and
$\Phi_3$~\cite{Petersen:2010cw}.
The dispersion of $\Psi_1-\Phi_1$ is qualitatively similar, but much
larger. Note that in our calculation, $\Psi_{1}$ is determined very
accurately for each event, and this dispersion cannot be attributed to
a finite event-plane resolution~\cite{Poskanzer:1998yz}.

This large dispersion shows that, although the initial geometry
specifies directed flow and $\Psi_1$
completely, the information on directed flow is not uniquely contained
in initial dipole asymmetry. In the language of
cumulants~\cite{Teaney:2010vd}, the dipole asymmetry
represents only the first term in an expansion, and higher
order terms contribute.
We have also checked that the distribution of $\Phi_1$  is flat: the
direction of the dipole asymmetry is uncorrelated with the reaction
plane (which is the $x$-axis in NeXSPheRIO), as expected~\cite{Teaney:2010vd}.

\begin{figure}[ht]
\includegraphics[width=\linewidth]{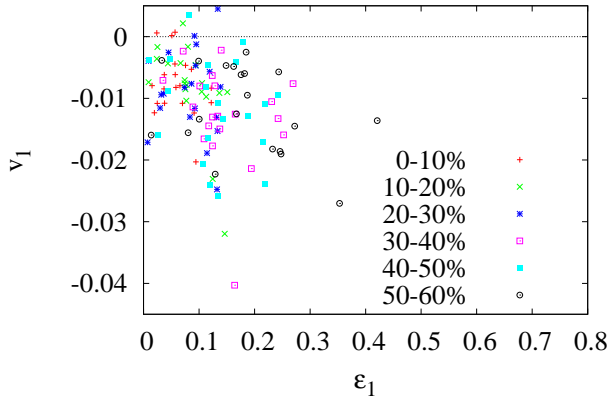}
\caption{(Color online)
Scatter plot of average $v_1$ in $0<p_t<1$~GeV/c, $|\eta|<1$ and initial dipole
asymmetry $\varepsilon_1$. Results for $v_1$ are corrected for net
transverse momentum.}
\label{fig:v1vseps1}
\end{figure}
Fig.~\ref{fig:v1vseps1} displays the values of $v_1$, averaged over
$0<p_t<1$~GeV/c, and the initial dipole asymmetry $\varepsilon_1$ for
the 120 hydro events used in our analysis.
This figure shows that there is no one-to-one
correspondence between $v_1$ and $\varepsilon_1$, though events
with a large dipole asymmetry give on average a larger $v_1$.

\section{Conclusions}
\label{s:conclusions}

We have computed the directed flow, $v_1$, created by initial state
fluctuations near midrapidity using event-by-event ideal hydrodynamics
for Au-Au collisions at RHIC. This $v_1$ is an even function of
(pseudo)rapidity and
has a specific $p_t$ dependence: it is negative below
$1$~GeV/c and positive above.
Its minimum value lies between $-0.02$ and $-0.01$ in the
centrality range $0-40$~\%.
This is in contrast to the $p_t$ dependence of the usual, rapidity-odd
directed flow, which is typically negative for all $p_t$ at SPS
(see the $v_1\{3\}$ results in Ref.~\cite{Alt:2003ab}) and
at RHIC~\cite{Adams:2005ca}.

The angle of directed flow $\Psi_1$ is correlated with the angle of
the initial dipole asymmetry $\Phi_1$, but with a large dispersion.
This confirms the idea of Teaney and Yan that the dipole asymmetry is
the mechanism creating $v_1$. However, it is only a rough picture.
$v_1$ is not proportional to $\varepsilon_1$ on an event-by-event
basis. More work is needed to understand how fluctuations in the
initial geometry are related to final momentum spectra.

Our ideal hydrodynamics results are in remarkable agreement with
preliminary experimental results~\cite{Luzum:2010fb,Agakishiev:2010ur}
inferred from STAR dihadron correlation data in mid-central Au-Au
collisions.
This quantitative agreement suggests that $v_1$ is less sensitive to
viscosity than $v_2$ and $v_3$. Detailed study of $v_1$ in viscous
hydrodynamics is left to future work.
Our results establish directed flow at midrapidity as a clear probe of
hydrodynamic behavior.
Experimental uncertainties on this $v_1$ are still large, but they can
be significantly reduced by carrying out dedicated analyses at RHIC
and LHC.

\begin{acknowledgments}

This work is funded by ``Agence Nationale de la Recherche'' under grant
ANR-08-BLAN-0093-01, by Cofecub under project Uc Ph 113/08;2007.1.875.43.9, by FAPESP under projects 09/50180-0, 09/16860-3 and 10/51479-6, and by CNPq under projects 305653/2007-5 and 301141/2010-0.

\end{acknowledgments}

\appendix
\section{Initial conditions for event-by-event hydrodynamics}
\label{s:app}

In event-by-event hydrodynamics, initial conditions are taken from a
microscopic model, which provides an energy-momentum tensor
$T^{\mu\nu}$. Since local thermal equilibrium is usually not achieved
in the microscopic model, $T^{\mu\nu}$ is not the energy-momentum
tensor of an ideal fluid.
By switching from the microscopic model to hydrodynamics, one must
modify $T^{\mu\nu}$ --- essentially approximating a yet-unknown thermalization or isotropisation mechanism.   In doing so, one must choose quantities to keep continuous during this transformation, while the rest of the energy-momentum tensor necessarily changes discontinuously.

A common procedure is to transform $T^{\mu\nu}$ into the local rest
frame~\cite{Werner:2010aa} where the momentum density $T^{0i}$
vanishes. The fluid velocity is then defined as the velocity of the
transformation, and the energy density is defined as $\epsilon\equiv
T^{00}$ in the local rest frame.
Mathematically, this boils down to a diagonalization of
$T^{\mu}_{\phantom{\mu}\nu}$~\cite{Hama:2004rr}: $u^\mu$ is the
normalized time-like eigenvector and $\epsilon$ is the corresponding
eigenvalue:
\begin{equation}
\label{eigenvalue}
T^{\mu}_{\phantom{\mu}\nu}u^\nu=\epsilon u^\mu.
\end{equation}
The pressure $P$ is then defined from $\epsilon$ using the equation of
state of the fluid, and the energy-momentum tensor of the fluid is
defined as usual as $T^{\mu\nu}_{\rm fluid}=(\epsilon+P)u^\mu u^\nu-Pg^{\mu\nu}$.   Thus the local rest frame, defined by the fluid velocity $u^\mu$, and the energy density in this frame are chosen to be continuous through the thermalization transformation.

By transforming back to a common lab frame, however, it becomes apparent that the
energy and momentum density are changed in this process, in an uncontrolled way.  For
simplicity, consider the case where
the transition between the microscopic model and hydrodynamics is
done at an initial time $t_0$ in the lab frame.  The total energy and momentum of the
fluid are the integrals of $T^{00}$ and $T^{0i}$ over the entire
space. The above procedure changes $T^{00}$ and $T^{0i}$, and
hence violates conservation of energy and momentum, even though the corresponding quantities in the local rest frame  $T_{\rm rest}^{0\mu} \equiv u_\nu T^{\mu\nu}$ are continuous.

Alternatively, rather than demanding continuity of $u^\mu$ and $\epsilon$,  one can instead choose a procedure that respects conservation of energy and momentum, which we describe here.

It should be noted that the above procedure is not always used~\cite{Petersen:2008dd}, and in hydrodynamic calculations without a model for the initial microscopic dynamics (e.g., when only the transverse profile is postulated from a model, while the initial fluid velocity is set by hand) the issue does not arise~\cite{Holopainen:2010gz,Schenke:2010rr}.


In general, the transition from the microscopic model to
hydrodynamics is done across a space-like ``freeze-in'' hypersurface $\Sigma$.
The total energy and momentum across $\Sigma$ is
\begin{equation}
P^\mu=\int_{\Sigma} T^{\mu\nu}d\sigma_\nu,
\end{equation}
where $d\sigma_\nu$ is the elementary time-like vector normal to
$\Sigma$.

Global conservation of energy and momentum demands continuity of $P^\mu$ across the freeze-in surface.  This can be ensured by demanding \textit{local} energy/momentum conservation ($\partial_\nu T^{\mu\nu}=0$), which requires continuity of the energy and momentum flux across the surface.
Thus, to generate initial conditions for ideal hydrodynamics in a way that respects energy-momentum conservation laws one must demand local continuity of $T^{\mu\nu}d\sigma_\nu$.   For example, in the case of a constant time surface, $d\sigma_\nu \propto (1,0,0,0)$, this amounts to continuity of the energy and momentum density $T^{0\mu}$  \cite{Petersen:2008dd}.

In general, this gives 4 continuity equations relating the hydrodynamic variables ($\epsilon$, $u^\mu$, $P$) to the energy-momentum tensor $T^{\mu\nu}$ from the microscopic model,
\begin{equation}
T^{\mu\nu}d\sigma_\nu = (\epsilon + P) u^\mu u^\nu  d\sigma_\nu - P g^{\mu\nu} d\sigma_\nu.
\end{equation}
Further specifying an equation state $P(\epsilon)$ uniquely determines the hydrodynamic initial conditions, without any additional freedom.

Note that this method agrees with the former method only at points along the freeze-in
surface where $d\sigma^\mu \propto u^\mu$, with $u^\mu$ defined by
Eq.~\eqref{eigenvalue}.  Everywhere else along the surface, demanding conservation of energy and momentum results in values for $\epsilon$ and $u^\mu$ in the hydrodynamic phase that are different from those acquired from the microscopic model by Eq.~\eqref{eigenvalue}.
%

\end{document}